\documentclass[12pt,eadjoint tfnpsf]{article}

\catcode`\@=11
\@addtoreset{equation}{section}

\global\arraycolsep=1pt
\usepackage{amsbsy,amssymb,latexsym,amsfonts}
\usepackage{mathrsfs}
\usepackage{graphicx}

\RequirePackage[dvips,usenames]{color}
\definecolor{fireblick}{rgb}{0.698039,0.133333,0.133333}

\newcommand{\beq}{\begin{equation}}
\newcommand{\eeq}{\end{equation}}
\newcommand{\bea}{\begin{eqnarray}}
\newcommand{\eea}{\end{eqnarray}}

\newcommand{\CD}{{\mathcal D}}
\newcommand{\CF}{{\mathcal F}}

\newcommand{\CN}{{\mathcal N}}
\newcommand{\CO}{{\mathcal O}}

\newcommand{\CW}{{\mathcal W}}

\def\Tr{\mathop{\rm Tr}}

\newcommand\diag{\mathrm{diag}}

\begin{document}
%
%
\begin{titlepage}

\begin{flushright}
\normalsize
~~~~
October, 2007 \\
OCU-PHYS 278 \\
\end{flushright}

\bigskip
\bigskip

\begin{center}
{\Large\bf Deformation of Dijkgraaf-Vafa Relation via \\
  Spontaneously Broken ${\mathcal N}=2$ Supersymmetry I\hspace{-.1em}I }
\end{center}

\bigskip

\begin{center}
{%
H. Itoyama$^{a,b}$\footnote{e-mail: itoyama@sci.osaka-cu.ac.jp}
\quad and \quad
K. Maruyoshi$^a$\footnote{e-mail: maruchan@sci.osaka-cu.ac.jp}
}
\end{center}

\bigskip

\begin{center}
$^a$ \it Department of Mathematics and Physics, Graduate School of Science\\
Osaka City University\\
\medskip

$^b$ \it Osaka City University Advanced Mathematical Institute (OCAMI)

\bigskip

3-3-138, Sugimoto, Sumiyoshi-ku, Osaka, 558-8585, Japan \\

\end{center}

\bigskip

\begin{abstract}
  We consider the matter induced part of the effective superpotential of $\CN=2$, $U(N)$ gauge model 
  in which $\CN=2$ supersymmetry is spontaneously broken to $\CN=1$,
  by using the properties of the chiral ring and the generalized Konishi anomaly equations
  derived in our previous paper arXiv:0704.1060.
  It is shown that the effective superpotential is related to the planar free energy of the matrix model
  by a formula which consists of two parts ---
  the well-known part due to Dijkgraaf-Vafa and the part that acts as a deformation of the couplings.
  These couplings are those of the original bare prepotential in the action and at the same time matrix model couplings.
\end{abstract}

\vfill

\setcounter{footnote}{0}
\renewcommand{\thefootnote}{\arabic{footnote}}

\end{titlepage}

\section{Introduction}
\label{sec:intro}
  In the last two decades, various investigations have been made 
  on the low energy effective action of supersymmetric gauge theory.
  It has been shown that the low energy effective action of $\CN=2$ supersymmetric gauge theory,
  which is governed by the effective prepotential, can be explicitly calculated, 
  by exploiting its powerful constraints associated with holomorphy \cite{SW} and instanton calculation \cite{Nekrasov}.
  In contrast to the fact that $\CN=2$ supersymmetric Yang-Mills theories are in the Coulomb phase, 
  $\CN=1$ supersymmetric gauge theories offer a wealth of vacua.
  Physically interesting phenomena, such as confinement and mass gap occur in low energy.
  It has been conjectured 
  in the context of the topological string theory and the gauge/gravity correspondence \cite{Vafa,CIV,CV}
  that the effective superpotential is related to the matrix model free energy \cite{DV}, 
  which we refer to as Dijkgraaf-Vafa relation.
  This relation has been shown in \cite{DGLVZ,CDSW,Ferrari1} by the purely field theoretical argument.
  (For subsequent developments on the calculus associated with the matrix model curve
  as algebraic integrable systems, see \cite{Chekhov1}.)
  
\begin{figure}[t]
    \begin{center}
    \includegraphics[scale=0.55]{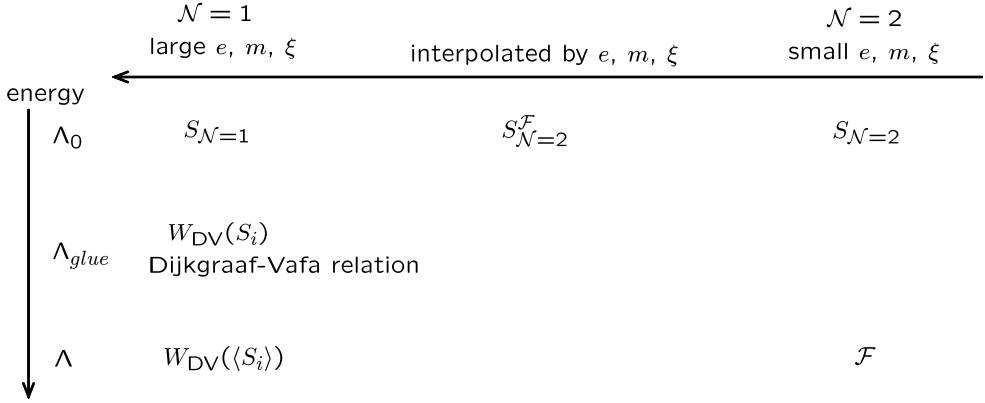}
    \caption{{\small Interpolation by Fayet-Iliopoulos parameters $e, m$ and $\xi$.
             At the energy $\Lambda_0$, 
             the action $S_{\CN=2}^\CF$ of the model \cite{FIS1,FIS2} reduces to the action $S_{\CN=2}$ in \cite{Gorsky}
             and the action $S_{\CN=1}$ in \cite{DV} in the small and large Fayet-Iliopoulos parameters limits
             respectively.}}
    \label{fig:inter}
    \end{center}
    \end{figure}

  More recently, a supersymmetric $U(N)$ gauge model, 
  in which $\CN=2$ supersymmetry is spontaneously broken to $\CN=1$, 
  has been found in \cite{FIS1,FIS2}, 
  and this model is the non-Abelian generalization of the Abelian model \cite{APT}.
  (See also \cite{FIS3,IMS} for the cases with hypermultiplet, 
  \cite{sugra} for $\CN=2$ supergravity and \cite{KMG} for related discussions.)
  It is not difficult to imagine that this model connects the above $\CN=2$ and $\CN=1$ theories.
  On the one hand, $\CN=2$ supersymmetry is restored in the small Fayet-Iliopoulos parameters limit.
  To be precise, in this limit, the action of the model \cite{FIS1,FIS2} reduces to 
  that of the extended $\CN=2$ supersymmetric Yang-Mills theory 
  whose effective superpotential has been discussed in the literature \cite{Gorsky,MN}.
  On the other hand, the action of the model reduces to that of the $\CN=1$ supersymmetric $U(N)$ theory
  with an adjoint chiral superfield $\Phi$ and a tree level superpotential $W(\Phi)$,
  which has been considered by \cite{DV,DGLVZ,CDSW},
  in the limit where the Fayet-Iliopoulos parameters are taken to be infinite \cite{Fujiwara}.
  Therefore, we can regard, at the classical level,
  the above two different theories as the particular limits of the model. 
  We illustrate this in Figure \ref{fig:inter}.
  
  So it is quite interesting to consider the quantum structure of this model:
  \textit{how is the effective superpotential? and how is the Dijkgraaf-Vafa relation deformed?}
  In \cite{IM}, we have started an analysis on the matter induced part of the effective superpotential of the model
  by computing the loop diagrams, following the spirit of \cite{DGLVZ}
  and have shown that the Dijkgraaf-Vafa relation is deformed 
  in the region of large Fayet-Iliopoulos parameters.
  We have determined the leading term of this deviation from the Dijkgraaf-Vafa relation.
  In this computation, however, we have to treat many interaction terms
  and it is technically difficult to calculate all the contributions to the effective superpotential.
  We have also derived a set of two generalized Konishi anomaly equations 
  on the two one-point functions $R(z)$ and $T(z)$.
  
  The aim of this paper is to obtain an exact expression which relates the effective superpotential 
  with the planar free energy of the matrix model.
  For this purpose, we use an alternative method 
  which is based on the properties of the chiral ring and the Konishi anomaly \cite{Konishi,CDSW}.
  In this approach, we do not need to take the Fayet-Iliopoulos parameters to be large.
  The effective superpotential consists of two parts both of which are written as operators 
  acting on the planar free energy of the bosonic one-matrix model.
  The first part is well-known from the case of \cite{DV} 
  while the second part acts as a (Whitham) deformation\footnote{See, for example, \cite{Gorsky}.} of the couplings.
  
  In \cite{Ferrari4}, the effective superpotential of a generic $\CN=1$ gauge model 
  containing the non-canonical gauge kinetic term has been derived, 
  so as to justify the important assumptions 
  of the matrix model and the generalized Konishi anomaly equations.
  The model we consider has been studied for a while \cite{FIS1,FIS2},
  as an non-Abelian generalization of \cite{APT}
  emphasizing the nature of partially and spontaneously broken $\CN=2$ supersymmetry, 
  and can be regarded as a distinguished class of a generic $\CN=1$ model.
  This paper is a sequel to our previous paper \cite{IM}, 
  where the generalized Konishi anomaly equations were already derived.
  
  The organization of this paper is as follows.
  In section \ref{sec:pre}, we review results of \cite{FIS1,FIS2,Fujiwara,IM}.
  Using the generalized Konishi anomaly equations \cite{IM}, 
  we obtain an explicit expression of the generating function of the one-point function $R(z)$ 
  in section \ref{sec:R}.
  Also, by making use of the solution of the generalized Konishi anomaly equation for the generating function $T(z)$, 
  we obtain the relation in section \ref{sec:effsuperpot}.
  Finally, we compare this result with the one derived from the diagrammatical computation \cite{IM}
  in section \ref{sec:comparison}.

\section{Preliminaries}
\label{sec:pre}
  In this section, we collect some known facts which are important for the analysis of this paper.
  In subsection \ref{sec:model}, we introduce the bare action of the model we study 
  and discuss the partial breaking of $\CN=2$ supersymmetry.
  In subsection \ref{sec:diagrammatical} and \ref{sec:konishi}, 
  we briefly review the results of our previous paper \cite{IM}.
  We explain the result from our diagrammatical computation in subsection \ref{sec:diagrammatical} and 
  derive the generalized Konishi anomaly equations in subsection \ref{sec:konishi}.
  
\subsection{The $U(N)$ gauged model with spontaneously broken $\CN=2$ supersymmetry}
\label{sec:model}
  The bare action of the model we study in this paper is
  \footnote{In \cite{FIS1, FIS2}, the action (\ref{S2}) is constructed, 
            following the gauging procedure of the general K\"ahler potential in \cite{WB},
            and restricting itself to be the one dictated by the special K\"ahler geometry.
            For the sake of completeness, we show the equivalence of (\ref{S2}) with the action in \cite{FIS1, FIS2} 
            in appendix A.}
    \bea
    S_{\CN=2}^{\CF}   
    &=&    \int d^4 x d^4 \theta 
           \left[
         - \frac{i}{2} {\rm Tr} 
           \left(  \bar{\Phi} e^{ad V} 
           \frac{\partial \CF(\Phi)}{\partial \Phi}
         - h.c.
           \right)
         + \xi V^0 
           \right] 
           \nonumber \\
    & &  + \left[
           \int d^4 x d^2 \theta
           \left(
         - \frac{i}{4} 
           \frac{\partial^2 \CF(\Phi)}{\partial \Phi^a \partial \Phi^b}
           \CW^{\alpha a} \CW^b_{\alpha}
         + e \Phi^0
         + m \frac{\partial \CF(\Phi)}{\partial \Phi^0}
           \right)
         + h.c.
           \right],      
           \label{S2}
    \eea
  where $V$ and $\Phi$ are the vector and chiral $\CN=1$ superfields
  whose on-shell components are ($A_\mu$, $\lambda^\alpha$) and ($\phi$, $\psi^\alpha$) respectively.
  In terms of $U(N)$ generators $t_a$, $a= 0, \ldots, N^2 -1$ ($a = 0$ refers to the overall $U(1)$ generator), 
  the superfield $\Psi = \{ V, \Phi \}$ is $\Psi = \Psi^a t_a$.
  (We normalize the generators as $\Tr (t_a t_b) = \delta_{ab}/2$.)
  Theoretical inputs are the electric and magnetic Fayet-Iliopoulos terms 
  which are two vectors or a rank two symmetric tensor in the isospin space 
  and are parameterized by the three real parameters $e, m, \xi$ in the $\CN=1$ superspace formalism we employ.
  In addition, the model contains an arbitrary input function $\CF(\Phi)$, 
  which we refer to as a bare prepotential.
  Its prototypical form is a single trace function of a polynomial in $\Phi$:
    \bea
    \CF(\Phi)
     =     \sum_{\ell=1}^{n+1} \frac{g_\ell}{(\ell+1)!} \Tr \Phi^{\ell+1},
           ~~
           {\rm deg} \CF = n+2.
           \label{prepot}
    \eea
    
  While this action is shown to be invariant under the $\CN=2$ supersymmetry
  transformations \cite{FIS1, FIS2},  the vacuum breaks half of 
  the $\CN=2$ supersymmetries.
  Extremizing the scalar potential, we obtain the condition 
    \bea
    \langle \frac{\partial^2 \CF(\Phi)}{\partial \Phi^0 \partial \Phi^0} \rangle 
     =   - \frac{e \pm i \xi}{m}.
           \label{Vcond}
    \eea
  The left hand side is a polynomial of order $n$ and determines the expectation value of the scalar field.
  In these vacua, the combination of the fermions, $(\lambda^\alpha \mp \psi^\alpha)/\sqrt{2}$, becomes massive, 
  while $(\lambda^\alpha \pm \psi^\alpha)/\sqrt{2}$ is massless, 
  whose overall $U(1)$ component is the Nambu-Goldstone fermion.
  In order to obtain the action on the vacua, we, therefore, have to redefine the superfields $V$ and $\Phi$ 
  such that the fermionic components of them mix as $\psi_{\pm}^\alpha = (\lambda^\alpha \pm \psi^\alpha)/\sqrt{2}$.
  In \cite{Fujiwara}, the action on the vacua has been obtained by taking this point into account
  and that the Fayet-Iliopoulos D-term can be included in the superpotential;
    \bea
    S_{\CN=1}^{\CF}   
    &=&    \int d^4 x d^4 \theta 
           \left[
         - \frac{i}{2} {\rm Tr} 
           \left(  \bar{\Phi} e^{ad V} 
           \frac{\partial \CF(\Phi)}{\partial \Phi}
         - h.c.
           \right) 
           \right] 
           \nonumber \\
    & &  + \left[
           \int d^4 x d^2 \theta
           \left(
         - \frac{i}{4} 
           \frac{\partial^2 \CF(\Phi)}{\partial \Phi^a \partial \Phi^b}
           \CW^{\alpha a} \CW^b_{\alpha}
         + W(\Phi)
           \right)
         + h.c.
           \right],      
           \label{S21}
    \eea
  where 
    \bea
    W(\Phi)
     =     \Tr \left[
           2 (e \pm i \xi) \Phi + m \sum_{\ell=1}^{n+1} \frac{g_\ell}{\ell!} \Phi^\ell
           \right]
           \label{Wtree}
    \eea
  is the single trace function of degree $n+1$ and $\CF(\Phi)$ is given by (\ref{prepot}).
  In (\ref{Wtree}), we have redefined $e, m, \xi$ such that they include the factor $1/\sqrt{2 N}$ 
  which comes from the overall $U(1)$ generator $t_0 = 1_{N \times N}/\sqrt{2 N}$.
  Also, it is understood that $V$ and $\Phi$ have been redefined as mentioned above.
  
  The action $S_{\CN=1}^{\CF}$ (\ref{S21}) is to be compared with
  that of the $\CN=1$, $U(N)$ gauge model with a single trace tree level superpotential $W(\Phi)$:
    \bea 
    S_{\CN=1}
     =     \int d^4 x d^4 \theta 
           \Tr \bar{\Phi} e^{ad V} \Phi
         + \left[
           \int d^4 x d^2 \theta
           \Tr \left(
           i \tau \CW^\alpha \CW_\alpha
         + W(\Phi)
           \right)
         + h.c. 
           \right],
           \label{S1}
    \eea
  where $\tau$ is a complex gauge coupling $\tau = \theta/2 \pi + 4 \pi i / g^2$.
  In \cite{FIS1}, it is checked that the second supersymmetry reduces 
  to the fermionic shift symmetry in the limit $m \rightarrow \infty$.
  The action $S_{\CN=1}^{\CF}$ in fact reduces to $S_{\CN=1}$
  in the limit $e, m, \xi \rightarrow \infty$ with $\tilde{g}_\ell \equiv m g_\ell$ ($\ell \geq 2$) 
  fixed \cite{Fujiwara}.
  We refer to this limit as $\CN=1$ limit.
  
  In this paper, we consider the matter-induced part of 
  the effective superpotential only by integrating out the massive degrees of freedom $\Phi$:
    \bea
    e^{i \int d^4 x (d^2 \theta W_{eff} + h.c. + ({\rm D-term}))}
     =     \int \CD \Phi \CD \bar{\Phi} e^{i S_{\CN=1}^{\CF}}.
           \label{Weff}
    \eea                                                          

\subsection{Diagrammatic analysis of the effective superpotential}
\label{sec:diagrammatical}
  Here, we review the diagrammatical computation of the effective superpotential \cite{IM}.
  For simplicity, we in this subsection consider the classical vacuum where $\langle \phi \rangle = 0$, 
  by setting the coupling constant as $m g_1 = - (e \pm i \xi)$.
  In this case, the unbroken gauge group is still $U(N)$.
  Also, we take $\mathcal{W}^\alpha$ (or $V$) as the background field 
    \footnote{The simplest background is that consisting of 
    a vanishing gauge field $A_\mu$ and a constant gaugino $\lambda^\alpha$, 
    which satisfies $\{ \lambda^\alpha, \lambda^\beta \} = 0$ \cite{AFH,IM}.
    This configuration implies that traces of more than two $\CW$ vanish.}.
  Therefore, the result of the diagrammatical computation can be written 
  in terms of the coupling constants $g_1$, $\tilde{g}_\ell$ ($\ell \geq 2$), the Fayet-Iliopoulos parameter $m$, 
  the glueball superfield $S \equiv - \Tr \CW^\alpha \CW_\alpha/64\pi^2 $ 
  and the overall $U(1)$ field strength $w^\alpha \equiv \Tr \CW^\alpha /8 \pi$.
  (The other Fayet-Iliopoulos parameters $e, \xi$ are always translated into $g_1$ and $m$
  by $m g_1 = - (e \pm i \xi)$.)
  
  Due to the diagrammatical computation, we can obtain the following formula \cite{IM}:
  the contribution from the $L$-loop diagrams which has $P$ propagators to the effective superpotential is, 
  up to terms including the overall $U(1)$ field strength $w^\alpha$, 
    \bea
    W_{eff}^{(L)}
     =     N \frac{\partial F_m^{(L)}}{\partial S}
         + W_2^{(L)}
         + W_3^{(L)},
           \label{WeffL}
    \eea
  where $W_2^{(L)}$ can be written as 
    \bea
    W_2^{(L)}
     =   - \frac{16 \pi^2 i \tilde{g}_3 P S}{m \tilde{g}_2 (L + 1)} 
           \left(
           \frac{\partial F_m^{(L)}}{\partial S}
           \right)
         + \hat{W}_2^{(L)}.
           \label{WeffL2}
    \eea
  In (\ref{WeffL2}), $\hat{W}_2^{(L)}$ is defined 
  by replacing, in the first term of r.h.s. of (\ref{WeffL}), one coupling constant according to
    \bea
    \tilde{g}_\ell
    \rightarrow
           \frac{16 \pi^2 i S}{N (L + 1)} g_{\ell + 1},
           ~~~~~~~{\rm for} ~~\ell = 3, \ldots, n.
           \label{Wefftype2}
    \eea
  and summing over all possibilities.
  Also, $W_3^{(L)}$ denotes the terms which include the higher order contributions in $1/m$.
  As discussed in \cite{IM}, $F_m^{(L)}$ in (\ref{WeffL}) can be identified with $L$-loop contribution
  to the planar free energy of the matrix model.
  Since $W_2^{(L)}$ and $W_3^{(L)}$ are $\CO(1/m)$,
  we can see that, in $\CN=1$ limit, we recover the result of \cite{DV, DGLVZ}.
  
  Although $W_2^{(L)}$ in (\ref{WeffL}) has been computed in \cite{IM}, 
  it is hard to obtain $W_3^{(L)}$ explicitly.
  In order to see this, we briefly recall some details of the computation.
  First of all, we start from (\ref{Weff}) and integrate $\bar{\Phi}$.
  This is easily done by setting the anti-holomorphic couplings $\bar{g}_\ell=0$ for $\ell \geq 3$.
  With this choice, the $\bar{\Phi}$-integral becomes a Gaussian integral 
  and we are left with the holomorphic part of the action
    \bea
    S_{\Phi}
    &=&    \int d^4 x d^2 \theta {\rm Tr}
           \Bigg[ 
           \sum_{\ell=2}^{n+1} \frac{\tilde{g}_\ell}{\ell!} \Phi^\ell  - \frac{i}{4}
           \sum_{\ell=3}^{n+1} \sum_{s=0}^{\ell-1} \frac{g_\ell}{\ell!} 
           (\CW^\alpha \Phi^s \CW_\alpha \Phi^{\ell-1-s})
           \nonumber \\
    & &    ~~~~~~~~~~~~~~~~~~~~~
         + \frac{1}{16 \bar{g}_2}
           \left( \bar{g}_1 \Phi - \frac{\partial \CF}{\partial \Phi} \right)
           \left(- \frac{2 m}{\nabla^2} + \frac{i}{4} \Phi \right)^{-1}
           \left( \bar{g}_1 \Phi - \frac{\partial \CF}{\partial \Phi} \right)
           \Bigg].
           \label{SPhi}
    \eea
  The first line is from $F$-term in $S_{\CN=1}^{\CF}$ 
  and the second line is due to the Gaussian integration of $\bar{\Phi}$.
  The latter can be expanded as
    \bea
    \frac{1}{16 \bar{g}_2}
    \left(
    \bar{g}_1 \Phi - \frac{\partial \CF}{\partial \Phi}
    \right)
    \left(
    - \frac{2 m}{\nabla^2} + \frac{i}{4} \Phi
    \right)^{-1}
    \left(
    \bar{g}_1 \Phi - \frac{\partial \CF}{\partial \Phi}
    \right)
     =     \frac{({\rm Im} g_1)^2}{8 \bar{\tilde{g}}_2} \Phi \nabla^2 \Phi
         + V(\Phi),
           \label{SPhi2}
    \eea
  where $V(\Phi)$ denotes the higher order interaction terms, which is not considered in \cite{IM}.
  Note that $V(\Phi)$ is $\CO(1/m)$.
  
  Secondly, we read off the Feynman rule from (\ref{SPhi}) and (\ref{SPhi2}).
  Collecting the quadratic terms we can determine the propagator.
  Because of the second term of (\ref{SPhi}) which does not exist in $S_{\CN=1}$, 
  the propagator is modified compared with that \cite{DGLVZ} of $S_{\CN=1}$.
  The higher order interaction terms in the first term in (\ref{SPhi}) are same as that \cite{DGLVZ} in $S_{\CN=1}$.
  On the other hand, the interaction terms in the second term in (\ref{SPhi}) do not exist in $S_{\CN=1}$.
  In addition, there are a lot of interaction terms in $V(\Phi)$.
  
  Finally, we compute the amplitude of the loop diagram.
  The amplitude of the non-planar diagram is exactly zero because of our choice of the background.
  (The detailed argument is found in \cite{IM}.)
  Therefore, we only have to consider the planar diagrams.
  From the contributions of the $L$-loop diagrams with $P$ propagators, we obtain (\ref{WeffL}) and (\ref{WeffL2}).
  The first term of (\ref{WeffL2}) is due to the fact that the propagator of the model is modified.
  Also, the second term of (\ref{WeffL2}) arises by considering the set of new vertices 
  which are seen in the first line of (\ref{SPhi}).
  The residual interaction $V(\Phi)$ is too complicated 
  to compute its contribution to the effective superpotential explicitly.
  We have denoted it as $W_3^{(L)}$ in (\ref{WeffL}).
  The result of the diagrammatical computation (\ref{WeffL}) is to be compared 
  with the effective superpotential which will be derived in section \ref{sec:effsuperpot},
  by making use of the generalized Konishi anomaly equations.
  Actually, as we will show in section \ref{sec:comparison}, $W_3^{(L)}$ exactly vanishes.
  
\subsection{Generalized Konishi anomaly equations}
\label{sec:konishi}
  An alternative approach to the effective superpotential 
  is to exploit and extend the properties of the $\CN = 1$ chiral ring and
  the generalized Konishi anomaly equations based on \cite{Konishi, CDSW}. 
  We will mainly use this approach in the rest of this paper.
  In this subsection, we derive the generalized Konishi anomaly equations 
  with respect to the chiral one-point functions \cite{IM}.
  
  The anomalous Ward identity of our model for the general transformation $\delta \Phi = f(\Phi, \CW)$ is 
    \bea 
         - \left< \frac{1}{64 \pi^2} 
           \left[ 
           \CW^\alpha , 
           \left[ \CW_\alpha , \frac{\partial f}{\partial \Phi_{ij}}
           \right]
           \right]_{ij}
           \right>   
     =     \left<
           {\rm Tr} f W'(\Phi)
           \right>
         - \left<
           \frac{i}{4} {\rm Tr} (f \CF'''(\Phi) \CW^\alpha \CW_\alpha)
           \right>,
           \label{WIforg}
    \eea
  in the chiral ring.
  The second term in r.h.s. is due to the fact 
  that the coefficient of $\CW^\alpha \CW_\alpha$-term in $S_{\CN=1}^\CF$ is function of $\Phi$, 
  rather than the constant $\tau$.
  Note that $W$ and $\CF$ are related as $W''(\Phi) = m \CF'''(\Phi)$.
  In terms of the two generating functions of the chiral one-point functions
    \bea
    R(z)
    &=&  - \frac{1}{64 \pi^2} \left<
           {\rm Tr} 
           \frac{\CW^\alpha \CW_\alpha}{z - \Phi}
           \right>,
           \nonumber \\
    T(z)   
    &=&    \left< {\rm Tr} 
           \frac{1}{z - \Phi}
           \right>,
    \eea
  the anomalous Ward identities (\ref{WIforg}) are
    \bea
    R(z)^2
    &=&    W'(z) R(z) + \frac{1}{4} f(z),
           \label{konishi1} \\
    2 R(z) T(z)
    &=&    W'(z) T(z) + 16 \pi^2 i \CF'''(z) R(z) + \frac{1}{4} c(z),
           \label{konishi2}
    \eea
  where $f(z)$ and $c(z)$ are polynomials of degree $n-1$ and 
    \bea
    \CF'''(z) 
     =     \sum_{\ell=2}^{n+1} \frac{g_\ell z^{\ell-2}}{(\ell - 2)!}
     =     \frac{W''(z)}{m}.
           \label{CF'''}
    \eea
  Since the explicit forms of $f(z)$ and $c(z)$ are not needed in the analysis of the subsequent sections, 
  we will not write it here.
  Note that the second term of r.h.s. of (\ref{WIforg}) does not contribute to the equation for $R(z)$ 
  because of the chiral ring relation ${\rm Tr} \CW^\alpha \CW_\alpha \CW^\beta \CW_\beta = 0$.
  The equation for $R(z) $ is, therefore, the same as that of \cite{CDSW}, 
  which is identified with the loop equation of the matrix model.
  On the other hand, the equation for $T(z)$ alters from that of \cite{CDSW}.
  This leads to the deformation of our effective superpotential from the well-known form 
  in the theory $S_{\CN=1}$ \cite{DV}.
  
\section{Solution of the anomaly equation for $R(z)$}
\label{sec:R}
  By solving the generalized Konishi anomaly equations (\ref{konishi1}) and (\ref{konishi2}), 
  we can obtain the explicit form of $R(z)$ and $T(z)$.
  In this section, we focus on $R(z)$.
  
  The classical vacua are determined by the condition (\ref{Vcond}) which is a polynomial of order $n$.
  If we denote the roots of (\ref{Vcond}) by $a_I$ ($I=1, \ldots n$), 
  the vacuum expectation value of the scalar field $\phi$ is
    \bea
    \langle \phi \rangle
     =     \diag (a_1, \ldots, a_1, a_2, \ldots a_2, \ldots, a_k \ldots, a_k).
           \label{vev}
    \eea
  Note that $k$ can be less than $n$.
  Let us denote the number of $a_I$ appearing in (\ref{vev}) by $N_I$.
  If $k < n$, corresponding $N_I$ ($I=k+1, \ldots, n$) are zero.
  We use indices $i,j$ ($i,j = 1,\ldots,k$) rather than $I,J$ when we refer only to nonvanishing $N_I$'s.
  In this notation, the gauge symmetry is broken to $\prod_{i=1}^k U(N_i)$ and $\sum_{i=1}^k N_i = N$.
  
  Let us first consider (\ref{konishi1}). 
  Its solution is
    \bea
    R(z)
     =     \frac{1}{2} \left( W'(z) - \sqrt{W'(z)^2 + f(z)} \right).
           \label{R}
    \eea
  The sign of square root is determined by the asymptotics $R(z) \sim S/z$ at large $z$.
  From the above form, we can see that $R(z)$ has cuts in the complex $z$ plane
  and is a meromorphic function on a Riemann surface $\Sigma$ of genus $n-1$
    \bea
    y^2
     =     W'(z)^2 + f(z).
    \eea
  Let us denote by $A_i$ $A$-cycles of $\Sigma$.
  In the semiclassical approximation where $f$ is small, to each cycle $A_i$ one can associate a zero of $W'$, $a_i$.
  Also, if we denote by $A_I$ ($I \neq i$) the contours which circle around $a_I$ with $I \neq i$, 
  these contours are trivial.
  Therefore, we have
    \bea
    S_i
     =     \oint_{A_i} R(z) d z
           ~~
           ({\rm for}~i = 1,\ldots,k),
           ~~~~
    0
     =     \oint_{A_I} R(z) d z
           ~~
           ({\rm for}~ I \neq i),
           \label{S}
    \eea
  where we have defined the contour integral to include a factor of $1/2 \pi i$.
  Also, we define $S = \sum_i S_i$.
  (\ref{S}) means that $y^2$ factorizes as
    \bea
    y^2
     =     W'(z)^2 + f(z)
     =     N_{n-k}(z)^2 F_{2k}(z).
           \label{fact}
    \eea
  $N_{n-k}(z)$ and $F_{2k}(z)$ are, respectively, polynomials of degree $n-k$ and $2k$.
  We obtain a reduced Riemann surface of genus $k-1$
    \bea
    y^2_{red}
     =     F_{2k}(z).
           \label{reduce}
    \eea
  
  Since $f(z)$ is a polynomial of degree $n-1$, \textit{a priori}, 
  $f(z)$ has $n$ undetermined coefficients.
  However, (\ref{fact}) produces $n-k$ constraints on the coefficients.
  Furthermore, the remaining undetermined coefficients are completely fixed by the first equation of (\ref{S}).
  Therefore, we can fix $y$ and $R(z)$ completely.
  
  For future reference, we consider the derivative of $R(z)$ with respect to $S_i$.
  From (\ref{R}), we obtain
    \bea
    \frac{\partial R(z)}{\partial S_i}
     =     \frac{\partial f(z)/\partial S_i}{4 \sqrt{W'(z)^2 + f(z)}}.
           \label{RS}
    \eea
  Also, by taking a derivative of (\ref{fact}), 
  we can see that $\partial f(z)/\partial S_i$ are proportional to $N_{n-k}$ 
  and therefore we can write $\partial f(z)/\partial S_i = N_{n-k}~ g_i(z)$ 
  where $g_i(z)$ are polynomials of degree $k-1$.
  Hence, (\ref{RS}) can be written as
    \bea
    \frac{\partial R(z)}{\partial S_i}
     =     \frac{g_i(z)}{4 F_{2k}(z)}
           ~~~
           ({\rm for}~i = 1,\ldots,k),
           \label{RS2}
    \eea
  where we have used the factorization condition (\ref{fact}) in the denominator.
  It is easy to see that $g_i(z)dz/4F_{2k}(z)$ ($i = 1,\ldots,k$) is a set of normalized holomorphic differentials
  on the reduced Riemann surface (\ref{reduce}).
  In fact, taking the derivative of (\ref{S}) with respect to $S_j$, we obtain
    \bea
    \delta_{ij}
     =     \oint_{A_i} \frac{g_j(z)}{4 F_{2k}(z)} dz.
    \eea
  Multiplying $N_j$ and summing over $j$, we obtain
    \bea
    N_i
     =     \oint_{A_i} \frac{\sum_j N_j g_j(z)}{4 F_{2k}(z)} dz.
           \label{Nigj}
    \eea
  
\section{Effective superpotential}
\label{sec:effsuperpot}
  In this section, we first state our formula for the effective superpotential and make a comment on this.
  In subsection \ref{sec:proof}, we provide a derivation of the formula.
  
  Let us define the one point functions as
    \bea
    v_\ell
     =   - \frac{1}{64 \pi^2} \langle \Tr \CW^\alpha \CW_\alpha \Phi^\ell \rangle,
           ~~~~
    u_\ell
     =     \langle \Tr \Phi^\ell \rangle,
           ~~~~
           {\rm for}~~ 1 \leq \ell \leq n+1.
    \eea
  In terms of $v_\ell$, we define $F$ as
    \bea
    \frac{\partial F}{\partial g_\ell}
     =     \frac{m}{\ell!} v_\ell,
           ~~~~
           {\rm for}~~ 1 \leq \ell \leq n+1.
           \label{Fg}
    \eea
  Since $v_\ell$ can be evaluated from $R(z)$ which has been fixed completely as we have seen in section \ref{sec:R}, 
  we can compute $F$ up to $g_\ell$-independent terms.
  Using $F$, the formula for the effective superpotential is given by
    \bea
    W_{eff}
     =      \sum_i N_i \frac{\partial F}{\partial S_i}
          + \frac{16 \pi^2 i}{m} \sum_{\ell = 2}^{n+1} g_\ell \frac{\partial F}{\partial g_{\ell - 1}},
            \label{weff}
    \eea
  up to $g_\ell$-independent terms.
  Indeed, the quantity $F$ can be identified with the free energy of the bosonic one matrix model 
  as we will see in section \ref{sec:comparison1}.
  Hence we find that $g_\ell$-dependent part of the effective superpotential of our model 
  can be obtained from the matrix model computation by the simple formula (\ref{weff}).
  In contrast to the case of $S_{\CN=1}$ \cite{DV}, we have the new term, the second term in (\ref{weff}).
  Because of its $1/m$ dependence
  (and since we can see in section \ref{sec:comparison} that $F$ depends only on $\tilde{g}_\ell$ and not on $m$), 
  the second term disappears in $\CN=1$ limit 
  where $m \rightarrow \infty$ with $\tilde{g}_\ell$ (for $\ell \geq 2$) fixed.
  Therefore, we obtain Dijkgraaf-Vafa formula as a particular limit of (\ref{weff}).
  
  In the theory $S_{\CN=1}$, it is known that 
  the full effective superpotential has the non-perturbative correction \cite{VY}
  which is called Veneziano-Yankielowicz term and do not depend on the coupling $g_\ell$.
  In \cite{DV}, it has been suggested 
  that the effective superpotential of the theory $S_{\CN=1}$ can be computed 
  from the matrix model including Veneziano-Yankielowicz term.
  The free energy of the matrix model in fact has $g_\ell$-independent term 
  by taking into account the volume of $U(\hat{N})$ group rotating the hermitian matrix $M$.
  From this term of the free energy, 
  we can obtain the well-known Veneziano-Yankielowicz term of the effective superpotential.
  
  In \cite{Ferrari4}, it has been shown 
  that the $g_\ell$-independent term is same as the well-known Veneziano-Yankielowicz term 
  using the instanton calculation \cite{MN}, for a generic $\CN=1$ gauge model.
  Here, however, we focus on only $g_\ell$-dependent part.
  
\subsection{Proof of the formula}
\label{sec:proof}
  Let us show the formula for the effective superpotential up to $g_\ell$-independent terms.
  To begin with, we take a derivative of (\ref{weff}) with respect to the coupling $g_\ell$, 
    \bea
    \frac{\partial W_{eff}}{\partial g_{\ell}}
    &=&    \frac{m}{\ell!} \sum_i N_i 
           \frac{\partial v_\ell}{\partial S_i}
         + \frac{16 \pi^2 i}{(\ell -1)!} v_{\ell-1}
         + \frac{16 \pi^2 i}{\ell!} \sum_{\ell'=2}^{n+1} g_{\ell'} 
           \frac{\partial v_\ell}{\partial g_{\ell'-1}}.
           \label{weff2}
    \eea
  Also, by taking a variational derivative of (\ref{Weff}) with respect to the coupling $g_\ell$, we obtain
    \bea
    \frac{\partial W_{eff}}{\partial g_\ell}
     =     \frac{m}{\ell!} u_\ell
         + \frac{16 \pi^2 i}{(\ell-1)!}v_{\ell-1}.
           \label{weff1}
    \eea
  By comparing (\ref{weff2}) and (\ref{weff1}), we obtain
    \bea
    u_\ell
     =     \sum_i N_i \frac{\partial v_\ell}{\partial S_i}
         + \frac{16 \pi^2 i}{m} \sum_{\ell'=2}^{n+1} g_{\ell'}
           \frac{\partial v_{\ell}}{\partial g_{\ell'-1}}.
    \eea
  Hence, once we prove the equation
    \bea
    T(z)
     =     \sum_i N_i \frac{\partial R(z)}{\partial S_i} 
         + \frac{16 \pi^2 i}{m} \sum_{\ell=2}^{n+1} g_{\ell} \frac{\partial R(z)}{\partial g_{\ell-1}}, 
           \label{T2}
    \eea
  the formula (\ref{weff}) follows as a truncation of (\ref{T2}) up to the first $n+1$ terms in the $1/z$ expansion.
  
  For this purpose, we start by solving the remaining generalized Konishi anomaly equation (\ref{konishi2}).
  By substituting (\ref{R}) into (\ref{konishi2}), we obtain
    \bea
    T(z)   
     =   - \frac{c(z)}{4 \sqrt{W'(z)^2 + f(z)}}
         + 8 \pi^2 i 
           \left(      
           \CF'''(z) 
         - \frac{W'(z) \CF'''(z)}{\sqrt{W'(z)^2 + f(z)}}     
           \right).
           \label{T}
    \eea
  Recall that $T(z)$ satisfies the following conditions;
    \bea
    N_i
     =     \oint_{A_i} T(z) dz,
           ~~~~
           {\rm for}~ i= 1,\ldots,k.
           \label{periodT}
    \eea
    
  Let us show that the right hand side of (\ref{T2}) is equal to the right hand side of (\ref{T}).
  As we have already observed in (\ref{RS2}), 
  $\frac{\partial R(z)}{\partial S_i} dz$ provides a set of normalized holomorphic differentials on the reduced curve.
  (\ref{periodT}) is, therefore, saturated by
    \bea
    \sum_{i} N_i \frac{\partial R(z)}{\partial S_i}
     =     \sum_i N_i \frac{g_i(z)}{4 F_{2k}(z)}
     \equiv
         - \frac{h(z)}{4 F_{2k}(z)},
    \eea
  with
    \bea
    N_i
     =  - \oint_{A_i} \frac{h(z)}{4 F_{2k}(z)} dz.
    \eea
  Introducing
    \bea
    D(z)
     \equiv
           c(z) - N_{n-k} h(z),
    \eea
  we obtain
    \begin{equation}
    0
     =     \oint_{A_I} 
           \left[
           \frac{- D(z)}{4 \sqrt{W'(z)^2 + f(z)}}
         + 8 \pi^2 i 
           \left(      
           \CF'''(z) 
         - \frac{W'(z) \CF'''(z)}{\sqrt{W'(z)^2 + f(z)}}     
           \right)
           \right]
           dz,
           ~~~
           1 \leq I \leq n.
           \label{eqd}
    \end{equation}
  
  On the other hand, the derivatives of $R(z)$ with respect to $g_{\ell}$ are
    \bea
    \frac{\partial R(z)}{\partial g_\ell}
     =     \frac{1}{2} 
           \left(
           \frac{\partial W'(z)}{\partial g_\ell} - \frac{W'(z) (\partial W'(z)/\partial g_\ell)}{\sqrt{W'(z)^2 + f(z)}}
           \right)
         - \frac{\partial f(z)/\partial g_\ell}{4 \sqrt{W'(z)^2 + f(z)}}.
           \label{Rg}
    \eea
  Recalling (\ref{CF'''}) as well as the definition of $W(z)$ and hence 
  $m \CF'''(z) = \sum_{\ell =1}^n g_{\ell + 1} \partial W'/\partial g_\ell$, 
  we obtain
    \bea
    \frac{16 \pi^2 i}{m} 
    \sum_{\ell = 1}^n g_{\ell + 1} \frac{\partial R(z)}{\partial g_\ell}
    &=&    8 \pi^2 i
           \left(      
           \CF'''(z) 
         - \frac{W'(z) \CF'''(z)}{\sqrt{W'(z)^2 + f(z)}}     
           \right)
           \nonumber \\
    & &    ~~~~~~~~~~~
         + \frac{16 \pi^2 i}{m}
           \left(
           \frac{- \sum_{\ell = 1}^n g_{\ell+1} \partial f(z)/\partial g_\ell}{4 \sqrt{W'(z)^2 + f(z)}}
           \right).
           \label{Rgg}
    \eea
  Our proof becomes complete as soon as we obtain
    \bea
    D(z)
     =     \frac{16 \pi^2 i}{m}
           \sum_{\ell = 1}^n g_{\ell+1} \frac{\partial f(z)}{\partial g_\ell}.
           \label{D}
    \eea
  Observing $0 = \partial S_I/\partial g_\ell = \frac{\partial}{\partial g_\ell} \oint_{A_I} R(z)$, 
  we obtain
    \bea
    0
    &=&    \oint_{A_I}
           \left[
           \frac{16 \pi^2 i}{m}
           \left(
           \frac{- \sum_{\ell = 1}^n g_{\ell+1} \partial f(z)/\partial g_\ell}{4 \sqrt{W'(z)^2 + f(z)}}
           \right)
         + 8 \pi^2 i
           \left(      
           \CF'''(z) 
         - \frac{W'(z) \CF'''(z)}{\sqrt{W'(z)^2 + f(z)}}     
           \right)
           \right]
           \nonumber \\
    & &    ~~~~~~~~~~~~~~~~~~~~~~~~~~~~~~~~~~~~~~~~~~~~~~~~~~~~~~~~~~~~~~~~~~~~~~~~~~~
           1 \leq I \leq n.
           \label{eqh}
    \eea
  Eq. (\ref{eqd}) and (\ref{eqh}) give
    \bea
    0
     =     \oint_{A_I}
           \frac{D(z) - \frac{16 \pi^2 i }{m}\sum_{\ell = 1}^n g_{\ell+1} \partial f(z)/\partial g_\ell}
           {4 \sqrt{W'(z)^2 + f(z)}} dz.
    \eea
  Expanding the integrand by a set of holomorphic differentials 
  $\{ z^\ell dz/\sqrt{W'(z)^2 + f(z)}, \ell = 0, \ldots n-1 \}$ of the original curve, we deduce (\ref{D}).
  
\section{Comparison with diagrammatical computation}
\label{sec:comparison}
  The effective superpotential (\ref{weff}) should be obtained from computing all the possible planar diagrams
  based on the procedure in the subsection \ref{sec:diagrammatical}.
  From (\ref{weff}), the $L$-loop contribution to the effective superpotential can be written as
    \bea
    W_{eff}^{(L)}
     =      \sum_i N_i \frac{\partial F^{(L)}}{\partial S_i}
          + \frac{16 \pi^2 i}{m} \sum_{\ell = 2}^{n+1} g_\ell \frac{\partial F^{(L)}}{\partial g_{\ell - 1}}.
            \label{weffL}
    \eea
  In this section, we compare this expression with the result of diagrammatical computation (\ref{WeffL}).
  At first sight, it seems that (\ref{weffL}) is different from (\ref{WeffL}):
  while the latter contains $W_3^{(L)}$ which contains in general higher order terms in $1/m$ in $\CN=1$ limit, 
  the former does not contain such terms.
  In section \ref{sec:comparison1}, 
  we will show that the first terms in two expressions (\ref{weffL}) and (\ref{WeffL}) are equal,
  which needs the consideration of the matrix model.
  Then, we show that the second term in (\ref{weffL}) are equivalent 
  to $W_2^{(L)}$ in (\ref{WeffL}) in section \ref{sec:comparison2}.
  This leads to that $W_3^{(L)}$ vanishes.
  
\subsection{Comparison with the matrix model}
\label{sec:comparison1}
  As discussed in \cite{IM}, 
  $F_m^{(L)}$ in (\ref{WeffL}) is the $L$-loop contribution to the free energy of the matrix model.
  Therefore, in this subsection, let us show that $F$ in (\ref{Fg}) or (\ref{weff}) is identified with 
  the free energy $F_m$ of the matrix model except for $g_\ell$-independent terms, 
  which leads to the identification $F^{(L)}$ in (\ref{weffL}) and $F_m^{(L)}$ in (\ref{WeffL}).
  The argument here is the same as that of \cite{CDSW}.
  
  The bosonic one matrix model is defined by integral of $\hat{N} \times \hat{N}$ hermitian matrix $M$.
  The definition of the free energy is
    \bea
    \exp \left( - \frac{\hat{N}^2}{g_m^2} F_m \right)
     =     \int d M \exp \left( - \frac{\hat{N}}{g_m} W(M) \right),
           \label{partition}
    \eea
  where
    \bea
    W(M)
     =     {\rm tr} \left[
           2 (e \pm i \xi) M + \sum_{\ell=1}^{n+1} \frac{\tilde{g}_\ell}{\ell!} M^\ell
           \right].
    \eea
  Note that the matrix size $\hat{N}$ is not related with the rank of the gauge group $N$.
  
  Let us define the matrix model resolvent as $ R_m (z) = \frac{g_m}{\hat{N}} \left< {\rm tr} \frac{1}{z - M} \right>$.
  With this, the loop equation reduces, in the planar limit, that is, the large $\hat{N}$ limit,
  to $ R_m(z)^2 = W'(z) R_m(z) + f_m(z)/4$
  whose form is the same as that of the generalized Konishi anomaly equation (\ref{konishi1}).
  A polynomial $f_m (z)$ is determined by the condition $g_m \hat{N}_i/\hat{N} = \oint_{A_i} dz R_m(z)$,
  where $\hat{N}_i$ is the number of the eigenvalues of $M$ near the $i$-th critical point 
  and each contour $A_i$ is defined to cycle the $i$-th critical point.
  If we identified the filling fraction $g_m \hat{N}_i/ \hat{N}$ with the glueball superfield $S_i$, 
  we can see that the polynomial $f_m(z)$ is equal to $f(z)$ in the gauge theory.
  Therefore, by the identification $S_i = g_m \hat{N}_i/ \hat{N}$, we can conclude $R_m(z) = R(z)$.
  
  As a final step, by taking a variational derivative of the partition function (\ref{partition}) 
  with respect to $g_\ell$, 
  we obtain
    \bea
    \frac{\partial F_m}{\partial g_\ell}
     =     \frac{g_m}{\hat{N}}
           \left< \frac{m}{\ell!} {\rm tr} M^\ell \right>
     =     \frac{m}{\ell!} v_\ell.
    \eea
  In the last equality, we have used $R_m(z) = R(z)$.
  This is the same equation as the definition of $F$ (\ref{Fg}).
  Hence, we conclude that $F$ in the effective superpotential (\ref{weff}) is the free energy of the matrix model
  up to $g_\ell$-independent terms.
  
\subsection{Comparison with the result of diagrammatical computation}
\label{sec:comparison2}
  In the last section, we have established the equivalence of $F^{(L)}$ and $F_m^{(L)}$.
  Here, we show the second term in (\ref{weffL}) is equal to $W_2^{(L)}$ in (\ref{WeffL}).
  
  Let us first consider the coupling dependence of the $F^{(L)} = F_m^{(L)}$.
  $F_m^{(L)}$ is the contribution from the $L$-loop diagrams to the matrix model free energy.
  From the form of the action of the matrix model (\ref{partition}), 
  we can read off the propagator which is proportional to $g_m/ \hat{N} \tilde{g}_2$ 
  and the vertices which are proportional to $\hat{N} \tilde{g}_\ell/ g_m$.
  Therefore, the amplitude of the $L$-loop diagrams with $P$ propagators and $V$ vertices is
    \bea
    \frac{f(\tilde{g}_3, \ldots, \tilde{g}_{n+1})}{\tilde{g}_2^P} g_m^{P-V} \hat{N}^{V-P+h},
    \eea
  where $f(\tilde{g}_3, \ldots, \tilde{g}_{n+1})$ is a function of $\tilde{g}_3, \ldots, \tilde{g}_{n+1}$ of degree $V$.
  $h$ is the number of the index loops and the factor $\hat{N}^h$ is due to the traces of the index loops.
  The function $f$ is determined by calculating 
  the symmetric factor and the coupling constants of each diagram we consider.
  Since we take the planar limit, the diagrams which should be considered have the topology of sphere 
  $\chi = V - P + h = 2$.
  By taking account of the factor in front of $F_m$ in (\ref{partition}), 
  we obtain the contribution of the $L$-loop planar diagrams
    \bea
    F_m^{(L)} 
     =     \frac{f(\tilde{g}_3, \ldots, \tilde{g}_{n+1})}{\tilde{g}_2^P} S^{h}.
           \label{F}
    \eea
  We have used the identification $g_m = S$ in the case of unbroken $U(N)$.
  Hence, if we use $L = h -1$, we have
    \bea
    N \frac{\partial F^{(L)}}{\partial S}
     =     N \frac{\partial F_m^{(L)}}{\partial S}
     =     N (L + 1) \frac{f(\tilde{g}_3, \ldots, \tilde{g}_{n+1})}{\tilde{g}_2^P} S^{L}.
           \label{FS}
    \eea
  
  Now, we are ready to show the second term in (\ref{weffL}) is $W_2^{(L)}$ in (\ref{WeffL}).
  From (\ref{WeffL}) and (\ref{WeffL2}), $W_2^{(L)}$ can be written as, 
    \bea
    W_2^{(L)}
     =     \frac{16 \pi^2 i S}{m (L + 1)}
           \left[
           (- P) \frac{\tilde{g}_3}{\tilde{g}_2} \frac{\partial F^{(L)}}{\partial S}
         + \frac{\partial F^{(L)}}{\partial S} |_{\tilde{g_\ell} \rightarrow \tilde{g}_{\ell + 1}}
           \right],
    \eea
  where $|_{\tilde{g_\ell} \rightarrow \tilde{g}_{\ell + 1}}$ means the procedure of
  changing the coupling constant by $\tilde{g}_\ell \rightarrow \tilde{g}_{\ell + 1}$ 
  for each coupling $\tilde{g}_\ell$ in $\partial F^{(L)}/ \partial S$
  and summing over all possibilities.
  The forms of (\ref{F}) and (\ref{FS}) lead to $\partial F^{(L)}/\partial S= (L+1)F^{(L)}/S$.
  Therefore, we derive 
    \bea
    W_2^{(L)}
     =     \frac{16 \pi^2 i}{m}
           \left[
           (- P) \frac{\tilde{g}_3}{\tilde{g}_2} F^{(L)}
         + F^{(L)}|_{\tilde{g_\ell} \rightarrow \tilde{g}_{\ell + 1}}
           \right]
    &=&    \frac{16 \pi^2 i}{m} \sum_{\ell = 2}^{n+1} g_\ell \frac{\partial F^{(L)}}{\partial g_{\ell - 1}}.
    \eea
  We have included $\ell =2$ term because $F_m^{(L)}$ do not contain $g_1$ 
  and thus $\partial F^{(L)}/\partial g_1 = 0$.

\section*{Acknowledgements}
  We thank Frank Ferrari, Alyosha Morozov and Futoshi Yagi for useful discussions.
  This work is supported in part by the Grant-in-Aid for Scientific Research (18540285) 
  from the Ministry of Education, Science and Culture, Japan.
  Support from the 21 century COE program ``Constitution of wide-angle mathematical basis focused on knots'' 
  is gratefully appreciated.
  
\appendix

\section*{Appendix}
\section{K\"ahler term in the action}
  In \cite{FIS1,FIS2}, the action has been constructed, 
  following the gauging procedure of the general K\"ahler potential in \cite{WB},
  restricting itself to be the one dictated by the special K\"ahler geometry.
  In this procedure, the action is \cite{FIS1,FIS2}
    \bea
    \frac{i}{2} (\Phi^a \bar{\CF}_a - \bar{\Phi}^a \CF_a) 
    + \int_0^1 d \alpha e^{\frac{i}{2} \alpha v^a (k_a - \bar{k}_a)} v^b \CD_b |_{v^a \rightarrow V^a},
    \label{FISaction}
    \eea
  where $\CF_a$ and $\bar{\CF}_a$ denote 
  $\partial \CF/\partial \Phi^a$ and $\partial \bar{\CF}/\partial \bar{\Phi}^a$ respectively.
  Also, $\partial_a = \partial/\partial \Phi^a$ and $\partial_{a^*} = \partial/\partial \bar{\Phi}^a$.
  In (\ref{FISaction}), $k_a$ are the Killing vectors and are generated by the Killing potentials $\CD_a$:
    \bea
    k_a
     =     k_a^{b~} \partial_b,
           ~~~
    k_a^{~b}
     =   - i g^{bc} \partial_{c^*} \CD_a,
           \label{killingvector}
    \eea
  which satisfies \cite{FIS1}
    \bea
    k_b^c \partial_c \Phi^a
     =     f^a_{bc} \Phi^c,
           ~~~
    k_b^c \partial_c \CF_a
     =   - f^a_{bc} \CF_c.
           \label{kandf}
    \eea
  Also, $\CD_a$ are given by
    \bea
    \CD_a
     =   - \frac{1}{2} (\CF_b f^b_{ac} \bar{\Phi}^c + \bar{\CF}_b f^b_{ac} \Phi^c).
           \label{killingpot}
    \eea
  
  At first sight, it seems that the form of (\ref{FISaction}) is different from the K\"ahler term
    \bea
    - \frac{i}{2} {\rm Tr} 
    \left(  \bar{\Phi} e^{ad V} 
    \frac{\partial \CF(\Phi)}{\partial \Phi}
    - h.c.
    \right)
    \label{IMaction}
    \eea
  in $S_{\CN=2}^{\CF}$ in subsection \ref{sec:model}.
  Let us show the equivalence of (\ref{FISaction}) and (\ref{IMaction}).
  Here, we work in Wess-Zumino gauge 
  and therefore we only have to show the equivalence of these up to second order in $V$.
  First of all, let us consider the zero-th order term in $V$.
  Using 
    \bea
    (t_a)_{ij} \frac{\partial \CF}{\partial \Phi_{ij}}
     =     (t_a)_{ij} \sum_{\ell = 1}^{n+1} \frac{g_\ell}{\ell!} (\Phi^\ell)_{ji}
     =     \sum_{\ell = 1}^{n+1} \frac{g_\ell}{\ell!} {\rm Tr} (t_a \Phi^\ell)
     =     \CF_a,
           \label{Fpartial}
    \eea
  where index $i=1, \ldots, N$ labels the fundamental representation, 
  the zero-th order term in (\ref{IMaction}) can be calculated as
    \bea
    \Tr \bar{\Phi} \frac{\partial \CF}{\partial \Phi}
     =     \bar{\Phi}_{ij} \frac{\partial \CF}{\partial \Phi_{ij}}
     =     \bar{\Phi}^a (t_a)_{ij} \frac{\partial \CF}{\partial \Phi_{ij}}
     =     \bar{\Phi}^a \CF_a.
    \eea
  Hence, the zero-th order terms in (\ref{FISaction}) and (\ref{IMaction}) are identical.
    
  Next, we turn to the linear term in $V$.
  The linear term in (\ref{FISaction}) is simply $V^a \CD_a$. 
  It is straightforward to observe
    \bea
    V^a \CD_a
    &=&   - \frac{1}{2} (V^a \CF_b f^b_{ac} \bar{\Phi}^c + h.c.)
            \nonumber \\
    &=&   - \frac{1}{2} 
            \left(
            V^a (t_a)_{ij} \frac{\partial \CF}{\partial \Phi_{ij}} f^b_{ac} \bar{\Phi}^c
          + h.c.
            \right)
            \nonumber \\
    &=&   - \frac{i}{2}
            \Tr  \left(
            \bar{\Phi} \left[ V, \frac{\partial \CF}{\partial \Phi} \right]
          - h.c.
            \right).
    \eea
  In the second equality, we have used (\ref{Fpartial}).
  This is the linear term in (\ref{IMaction}).
  
  Finally, let us consider the $V^2$ term in (\ref{FISaction}).
  By using (\ref{killingvector}) and (\ref{killingpot}), we derive
    \bea
    \frac{i}{4} V^a V^b (k_a - \bar{k}_a) \CD_b
    &=&    \frac{1}{2} V^a V^b g^{cd} (\partial_{d^*} \CD_a) (\partial_c \CD_b)
           \nonumber \\
    &=&  - \frac{1}{4} V^a V^b g^{cd} (\partial_{d^*} \CD_a) (\CF_{ec} f^e_{bf} \bar{\Phi}^f + \bar{\CF}_e f^e_{bc}).
           \label{2nd}
    \eea
  Since $g^{cd} (\partial_{d^*} \CD_a) \CF_{ec} = i k_a^c \partial_c \CF_e = - i f^c_{ae} \CF_c$ by (\ref{kandf}), 
  the first term can be written as
    \bea
    \frac{i}{4} V^a V^b f^c_{ae} \CF_c f^e_{bf} \bar{\Phi}^f.
    \eea
  On the other hand, by using the formula (we will show this formula below)
    \bea
    \CF_a f^a_{bc} 
     =   - \CF_{ac} f^a_{bd} \Phi^d,
           \label{formula1}
    \eea
  we can compute the second term of (\ref{2nd}) as follows:
    \bea
    - \frac{1}{4} V^a V^b g^{cd} (\partial_{d^*} \CD_a) \bar{\CF}_e f^e_{bc}
    &=&    \frac{1}{8} V^a V^b g^{cd} 
           ( \CF_f f^f_{ad} \bar{\CF}_e f^e_{bc} 
         + \bar{\CF}_{fd} f^f_{ag} \Phi^g \bar{\CF}_e f^e_{bc} )
           \nonumber \\
    &=&  - \frac{i}{4} V^a V^b f^c_{ag} \Phi^g \bar{\CF}_e f^e_{bc}.
    \eea
  Therefore, (\ref{2nd}) is
    \bea
    \frac{i}{4} (V^a V^b f^c_{ae} \CF_c f^e_{bf} \bar{\Phi}^f - h.c.)
     =   - \frac{i}{4} \Tr 
           \left(
           \bar{\Phi} \left[V, \left[V, \frac{\partial \CF}{\partial \Phi} \right] \right] - h.c.
           \right),
    \eea
  which proves the equivalence of the $V^2$ terms in (\ref{FISaction}) and (\ref{IMaction}).
  
  Let us show the formula (\ref{formula1}). 
  From (\ref{killingvector}) the first equation in (\ref{kandf}), we can write
    \bea
    f^a_{bc} \Phi^c
     =   - i g^{ac} \partial_{c^*} \CD_b
     =     \frac{i}{2} g^{ac} (\CF_d f^d_{bc} + \bar{\CF}_{dc} f^d_{be} \Phi^e).
    \eea
  Multiplying $g_{ha}$ by the above equation, 
    \bea
    g_{ha} f^a_{bc} \Phi^c
     =     \frac{i}{2} \CF_d f^d_{bh} + \frac{i}{2} \bar{\CF}_{dh} f^d_{be} \Phi^e
     =     \frac{i}{2} \CF_d f^d_{bh} + g_{dh} f^d_{be} \Phi^e + \frac{i}{2} \CF_{dh} f^d_{be} \Phi^e.
    \eea
  Therefore, we have shown the formula.



\begin{thebibliography}{99}
  
\bibitem{SW}
  N.~Seiberg and E.~Witten,
  Nucl.\ Phys.\  B {\bf 426} (1994) 19
  [Erratum-ibid.\  B {\bf 430} (1994) 485]
  [arXiv:hep-th/9407087];
%
  Nucl.\ Phys.\  B {\bf 431} (1994) 484
  [arXiv:hep-th/9408099].
  
\bibitem{Nekrasov}
  N.~A.~Nekrasov,
  Adv.\ Theor.\ Math.\ Phys.\  {\bf 7} (2004) 831
  [arXiv:hep-th/0206161];

  N.~Nekrasov and A.~Okounkov,
  arXiv:hep-th/0306238.

\bibitem{Vafa}
  C.~Vafa,
  J.\ Math.\ Phys.\  {\bf 42} (2001) 2798
  [arXiv:hep-th/0008142].

\bibitem{CIV}
  F.~Cachazo, K.~A.~Intriligator and C.~Vafa,
  Nucl.\ Phys.\  B {\bf 603} (2001) 3
  [arXiv:hep-th/0103067].
  
\bibitem{CV}
  F.~Cachazo and C.~Vafa,
  [arXiv:hep-th/0206017].

\bibitem{DV}
  R.~Dijkgraaf and C.~Vafa,
  Nucl.\ Phys.\  B {\bf 644} (2002) 3
  [arXiv:hep-th/0206255];
%
  Nucl.\ Phys.\  B {\bf 644} (2002) 21
  [arXiv:hep-th/0207106];
%
  [arXiv:hep-th/0208048].
  
\bibitem{DGLVZ}
  R.~Dijkgraaf, M.~T.~Grisaru, C.~S.~Lam, C.~Vafa and D.~Zanon,
  Phys.\ Lett.\  B {\bf 573} (2003) 138
  [arXiv:hep-th/0211017].
  
\bibitem{CDSW}
  F.~Cachazo, M.~R.~Douglas, N.~Seiberg and E.~Witten,
  JHEP {\bf 0212} (2002) 071
  [arXiv:hep-th/0211170].

\bibitem{Ferrari1}
  F.~Ferrari,
  JHEP {\bf 0606} (2006) 039
  [arXiv:hep-th/0602249];
%
  Nucl.\ Phys.\  B {\bf 770} (2007) 371
  [arXiv:hep-th/0701220];

  F.~Ferrari and V.~Wens,
  arXiv:0710.2978 [hep-th].
  
\bibitem{Chekhov1}
  L.~Chekhov and A.~Mironov,
  Phys.\ Lett.\  B {\bf 552} (2003) 293
  [arXiv:hep-th/0209085];
  
  S.~G.~Naculich, H.~J.~Schnitzer and N.~Wyllard,
  Nucl.\ Phys.\  B {\bf 651} (2003) 106
  [arXiv:hep-th/0211123];
  
  V.~A.~Kazakov and A.~Marshakov,
  J.\ Phys.\ A  {\bf 36} (2003) 3107
  [arXiv:hep-th/0211236];
  
  H.~Itoyama and A.~Morozov,
  Nucl.\ Phys.\  B {\bf 657} (2003) 53
  [arXiv:hep-th/0211245];
  
  S.~G.~Naculich, H.~J.~Schnitzer and N.~Wyllard,
  JHEP {\bf 0301} (2003) 015
  [arXiv:hep-th/0211254];

  H.~Itoyama and A.~Morozov,
  Phys.\ Lett.\  B {\bf 555} (2003) 287
  [arXiv:hep-th/0211259];
%
  Prog.\ Theor.\ Phys.\  {\bf 109} (2003) 433
  [arXiv:hep-th/0212032];

  L.~Chekhov, A.~Marshakov, A.~Mironov and D.~Vasiliev,
  Phys.\ Lett.\  B {\bf 562} (2003) 323
  [arXiv:hep-th/0301071];
  
  A.~Dymarsky and V.~Pestun,
  Phys.\ Rev.\  D {\bf 67} (2003) 125001
  [arXiv:hep-th/0301135];
  
  H.~Itoyama and A.~Morozov,
  Int.\ J.\ Mod.\ Phys.\  A {\bf 18} (2003) 5889
  [arXiv:hep-th/0301136];
  
H.~Itoyama and H.~Kanno,
  Phys.\ Lett.\  B {\bf 573} (2003) 227
  [arXiv:hep-th/0304184];

  S.~Aoyama and T.~Masuda,
  JHEP {\bf 0403} (2004) 072
  [arXiv:hep-th/0309232];

  A.~Alexandrov, A.~Mironov and A.~Morozov,
  Int.\ J.\ Mod.\ Phys.\  A {\bf 19} (2004) 4127
  [Teor.\ Mat.\ Fiz.\  {\bf 142} (2005) 419]
  [arXiv:hep-th/0310113];

  H.~Itoyama and H.~Kanno,
  Nucl.\ Phys.\  B {\bf 686} (2004) 155
  [arXiv:hep-th/0312306];
  
  E.~Konishi,
  arXiv:0707.0387 [hep-th].
  
\bibitem{FIS1}
  K.~Fujiwara, H.~Itoyama and M.~Sakaguchi,
  Prog.\ Theor.\ Phys.\  {\bf 113} (2005) 429
  [arXiv:hep-th/0409060];
  [arXiv:hep-th/0410132].

\bibitem{FIS2}
  K.~Fujiwara, H.~Itoyama and M.~Sakaguchi,
  Nucl.\ Phys.\  B {\bf 723} (2005) 33
  [arXiv:hep-th/0503113].
  
\bibitem{APT}
  I.~Antoniadis, H.~Partouche and T.R.~Taylor, 
  Phys.\ Lett.\ B {\bf 372} (1996) 83, [arXiv:hep-th/9512006].

\bibitem{FIS3}
  K.~Fujiwara, H.~Itoyama and M.~Sakaguchi,
  Nucl.\ Phys.\  B {\bf 740} (2006) 58
  [arXiv:hep-th/0510255];
  Prog.\ Theor.\ Phys.\ Suppl.\  {\bf 164} (2007) 125
  [arXiv:hep-th/0602267];
  AIP Conf.\ Proc.\  {\bf 903} (2007) 521
  [arXiv:hep-th/0611284].
  
\bibitem{IMS}
  H.~Itoyama, K.~Maruyoshi and M.~Sakaguchi,
  arXiv:0709.3166 [hep-th].

\bibitem{sugra}
  S.~Ferrara, L.~Girardello and M.~Porrati,
  Phys.\ Lett.\  B {\bf 366} (1996) 155
  [arXiv:hep-th/9510074];
  
  P.~Fre, L.~Girardello, I.~Pesando and M.~Trigiante,
  Nucl.\ Phys.\  B {\bf 493} (1997) 231
  [arXiv:hep-th/9607032];
  
  J. Louis, 
  arXiv:hep-th/0203138;
  
  H.~Itoyama and K.~Maruyoshi,
  Int.\ J.\ Mod.\ Phys.\  A {\bf 21} (2006) 6191
  [arXiv:hep-th/0603180];
  
  K.~Maruyoshi, 
  arXiv:hep-th/0607047.
  
\bibitem{KMG}
  P.~Kaste and H.~Partouche,
  JHEP {\bf 0411} (2004) 033
  [arXiv:hep-th/0409303];
  
  P.~Merlatti,
  Nucl.\ Phys.\  B {\bf 744} (2006) 207
  [arXiv:hep-th/0511280];
  
  L.~Girardello, A.~Mariotti and G.~Tartaglino-Mazzucchelli,
  JHEP {\bf 0603} (2006) 104
  [arXiv:hep-th/0601078].
  
\bibitem{Gorsky}
  A.~Gorsky, I.~Krichever, A.~Marshakov, A.~Mironov and A.~Morozov,
  Phys.\ Lett.\  B {\bf 355} (1995) 466
  [arXiv:hep-th/9505035];
  
  T.~Nakatsu and K.~Takasaki,
  Mod.\ Phys.\ Lett.\  A {\bf 11} (1996) 157
  [arXiv:hep-th/9509162];

  H.~Itoyama and A.~Morozov,
  Nucl.\ Phys.\  B {\bf 477} (1996) 855
  [arXiv:hep-th/9511126];
%
  Nucl.\ Phys.\  B {\bf 491} (1997) 529
  [arXiv:hep-th/9512161];

  A.~Gorsky, A.~Marshakov, A.~Mironov and A.~Morozov,
  Nucl.\ Phys.\  B {\bf 527} (1998) 690
  [arXiv:hep-th/9802007];

  J.~D.~Edelstein, M.~Marino and J.~Mas,
  Nucl.\ Phys.\  B {\bf 541} (1999) 671
  [arXiv:hep-th/9805172];
  
  J.~D.~Edelstein, M.~Gomez-Reino, M.~Marino and J.~Mas,
  Nucl.\ Phys.\  B {\bf 574} (2000) 587
  [arXiv:hep-th/9911115];

  K.~Takasaki,
  Prog.\ Theor.\ Phys.\ Suppl.\  {\bf 135} (1999) 53
  [arXiv:hep-th/9905224];

\bibitem{MN}
  A.~Marshakov and N.~Nekrasov,
  JHEP {\bf 0701} (2007) 104
  [arXiv:hep-th/0612019].
  
\bibitem{Fujiwara}
  K.~Fujiwara,
  Nucl.\ Phys.\ B {\bf 770} (2007) 145
  [arXiv:hep-th/0609039].

\bibitem{IM}
  H.~Itoyama and K.~Maruyoshi,
  Phys.\ Lett.\  B {\bf 650} (2007) 298
  [arXiv:0704.1060 [hep-th]];
  
  K.~Maruyoshi,
  arXiv:0710.2154 [hep-th].
  
\bibitem{Konishi}
  K.~Konishi,
  Phys.\ Lett.\  B {\bf 135} (1984) 439.
  
\bibitem{Ferrari4}
  F.~Ferrari,
  arXiv:0709.0472 [hep-th].
  
\bibitem{WB}
  J.~Wess and J.~Bagger,
{\it  Princeton, USA: Univ. Pr. (1992) 259 p}.

\bibitem{AFH}
  R.~Argurio, G.~Ferretti and R.~Heise,
  Int.\ J.\ Mod.\ Phys.\  A {\bf 19} (2004) 2015
  [arXiv:hep-th/0311066].

\bibitem{VY}
  G.~Veneziano and S.~Yankielowicz,
  Phys.\ Lett.\  B {\bf 113} (1982) 231.
  
\end{thebibliography}
\end{document}